\documentstyle[times,pramana,epsfig,floats]{ias}

\newcommand{\ba}{\begin{array}}
\newcommand{\ea}{\end{array}}
\newcommand{\bd}{\begin{displaymath}}
\newcommand{\ed}{\end{displaymath}}
\newcommand{\be}{\begin{equation}}
\newcommand{\ee}{\end{equation}}
\newcommand{\bea}{\begin{eqnarray}}
\newcommand{\eea}{\end{eqnarray}}

\def\q2 {q^2}

\begin{document}

\title{Hybrid Stars}
\author{Ashok Goyal}
\address{Department of Physics \& Astrophysics \\
University of Delhi, Delhi - 110 007.}

\abstract{
Recently there have been important developments in the determination
of neutron star masses which put severe constraints on the composition
and equation of state (EOS) of the neutron star matter. Here we study the
effect of quark and nuclear matter mixed phase on mass
radius relationship  of neutron stars employing recent models from two
classes of EOS's and discuss their implications.}

\maketitle

It is generally believed that the evolutionary journey of a star after
it has exhausted all its fuel culminates into the formation of a
compact object in the form of a {\bf white dwarf}, a {\bf neutron star} or a
{\bf  black hole} depending on its mass. White dwarfs support themselves
 against
gravitational collapse by the electron degeneracy pressure, neutron
stars largely by the pressure of degenerate neutrons whereas, the
black holes are completely collapsed stars. Neutron stars are the most
compact and dense stars known. They typically compress a solar
mass matter into a tiny radius of 10 Km with densities in the
core reaching several times the nuclear density. With such densities in
the core, they themselves can take various forms, for example they
could be composed of normal nuclear matter with hyperons and/or
condensed mesons. The matter at such densities may undergo phase
transition to constituent quark matter. It would then be energetically
profitable for the u-d matter to convert itself into u-d-s matter
through weak interactions thereby lowering its energy per baryon. If
it so happens that the energy per baryon of such matter called {\bf
strange matter} is the true 
ground state of matter with energy per baryon less
than that of iron, the most stable nuclei, the whole star
will convert itself into a {\bf strange star } with
vastly different characteristics. In case strange matter is not the
true ground state, the neutron star may have a quark core followed by
a mixed phase and nuclear mantle at the top. Such stars are called
{\bf hybrid stars }. A neutron star comes into existence through a
cataclysmic process at time scale large compared to not only the strong and
electromagnetic interaction time scales but weak interaction
scale as well. Further the electrostatic repulsion being so much
stronger compared to gravitational attraction, the matter is
electrically neutral and typical Fermi momenta of constituents being
large compared to its temperature, neutron star is composed of {\bf
cold, degenerate, charge neutral matter in $\beta$-equilibrium}. At
densities  $< 2 \times 10^{-3}\rho_0$  ($\rho_0=0.16~ nucleons~/fm^3~$ is
the equilibrium density of charged nuclear matter in nuclei), the
matter is assumed to be in the form of a Coulomb lattice (to minimize
energy) of nuclei immersed in a relativistic degenerate electron gas. In the
lower part of the density range $10^{-3}\rho_0~<~\rho~<10\rho_0~$ the
neutrons leak
 out of the nuclei and a Fermi liquid
of neutrons, protons and electrons start building up. At higher
densities there are several possibilities including the occurance of
muons, condensation of negatively charged pions and kaons, appearance
of hyperons and finally, the transition to quark matter.

Matter at such high densities has not been produced in the laboratory
and there is no available data on nuclear matter interactions. The
quantities of interest are the phases and composition of neutron star
matter, its energy density and pressure which determine the equation
of state (EOS). The EOS upto nuclear densities is fairly well
accounted for on the basis of measured nuclear data and nucleonic
interactions. Above nuclear densities there are basically two
approaches: one is to take the interaction between constituents from
realistic fitting with known scattering data and then use the
techniques of many body theory to calculate correlations. The other is
is to take a relativistic mean field type of model with couplings
treated as parameters to fit observable quantities. Both approaches
suffer from lack of experimental data. Whereas many body approach is
well understood, two nucleon interactions are fairly well known,
higher body interactions are not well characterised and the approach
is non-relativistic. Mean field theoretical models can easily
incorporate many constituents, but the complicated correlations are
simplified in terms of vacuum expectation of mean fields which are
fitted to insufficient data.

As discussed above, if there is a phase transition to quark matter,
the entire star may convert itself into a strange star or a hybrid
star depending on whether the strange matter is the true ground state
of matter or not.. In earlier studies the phase tramsition
was characterised as a first order transition with a single component
viz baryon number and charge neutrality was strictly enforced
in each phase separately. This gave rise to constant pressure
 (liquid-vapour) type phase transition and since in a star, the
pressure increases monotonically with density as we go from the 
surface to the core, mixed phase was strictly prohibited. It was 
pointed out by Glendenning \cite{1} that matter in neutron star has two components
,namely the conserved baryon number and the electric charge, therefore
the correct application of Gibb's phase rule is that the chemical
potential corresponding to baryon number and charge conservation
ie. $\mu_B$ and $\mu_Q$ , the temperature and the pressure in two 
phases are equal ie.
$$\mu_B(h) = \mu_B(q) \quad \quad ; \quad \quad \mu_Q(h) = \mu_Q(q)$$
$$ p_h(\mu_B,\mu_Q,T) = p_q(\mu_B,\mu_Q,T)$$
and charge neutrality only demands Global conservation
$$ \chi Q_q(\mu_B,\mu_Q,T) + (1-\chi)Q_h(\mu_B,\mu_Q,T) =0$$
where $\chi$ is the fraction of the volume occupied by the quark phase. 
The freedom available to the system to rearrange concentration of charges
 for a given fraction of phases $\chi$, results in variation of the pressure
 through the mixed phase. 
 The structure of the star in hydrostatic equilibrium is determined by
 solving the 
Tolman-Openheimer-Volkoff equations
and the only information required  is the knowledge of the EOS. For
the EOS we use a RFT model taken 
from Glendenning 
and a potential model incorporating relativistic corrections and three
body interactions given by Akaml, Pandharipande and Ravenhall \cite{2}
. Both classes of models admit 
 mixed quark-nucleon phase in the core \cite{3}. In fig
1 we see that the effect of the existence of 
mixed quark-nuclear matter phase in the core of neutron stars is to reduce the 
maximum mass. The effect is more pronounced for the relativistic mean
field theory model than for the potential model. 
In contrast, a strange star mass varies
as $\sim R^3$ for $M \sim 0.5 M_0$ and gravity plays minimal role, bag
pressure provides the confinement and the star is self bound. As mass
increases gravity becomes important and the star reaches a maximum
mass and the strange star radii decrease with decrease in
mass and there is no minimum mass.

During the last thirty years since the discovery of first pulsar in
1968 by Hewish et.al. close to two thousand pulsars using radio
telescope and X-ray probes in space have been discovered in a variety
of circumstances; i) as isolated radio sources at times in binary
orbits with other stars, ii) as X-ray pulsars and X-ray bursters in
X-ray binary system and iii) most recently by Rossi XTE X-ray
satellite as KHz quasi periodic oscillations and burst oscillations in
low mass X-ray binaries. This has led not only to their identification
as neutron stars but has provided physical laboratory with
unprecedented potentialities to perform tests of GTR and to obtain
information on the EOS of high density matter its composition and
phase transition.
 Masses of ten neutron stars have been measured by
observing relativistic effects in binary orbits and all of them have
masses in the range $1.35 +0.04 M_0$ \cite{4} 
with exceptions of PSR J1012 of
mass $2.1 + 0.4 M_0$. Masses of X-ray pulsars are meadured less
accurately and recent observations for Vela X-1 and Cygnus X-2 give
$1.9 + 0.2 M_0$ and $1.8 + 0.4 M_0$ respectively \cite{5}. The QPO observations 
correspond to masses $ \sim 2 M_0$ \cite{6}.
Thus if large masses of neutron stars are confirmed and complemented by
other neutron star masses $\sim 2 M_0$,
i) EOS is severely restricted and
only stiff EOS's without any significant phase transition below $5n_0$
are allowed.
ii) On the other hand if heavy neutron stars prove erroneous
by more detailed observations and masses like those of binary pulsars
($\sim 1.4 M_0$), this will indicate that accretion does not produce
heavier stars which will mean either a soft EOS or significant phase
transition at few times the nuclear saturation density. 
iii) Observation of
a star of mass $\leq M_0$ and radius $\leq 10 Km$ or a submilli-second
pulsar would be a good strange star candidate.

\noindent {\bf Acknowledgements :}   This work was supported in part by UGC and
SERC scheme of DST.  

\begin{figure}[t]
\epsfig{file=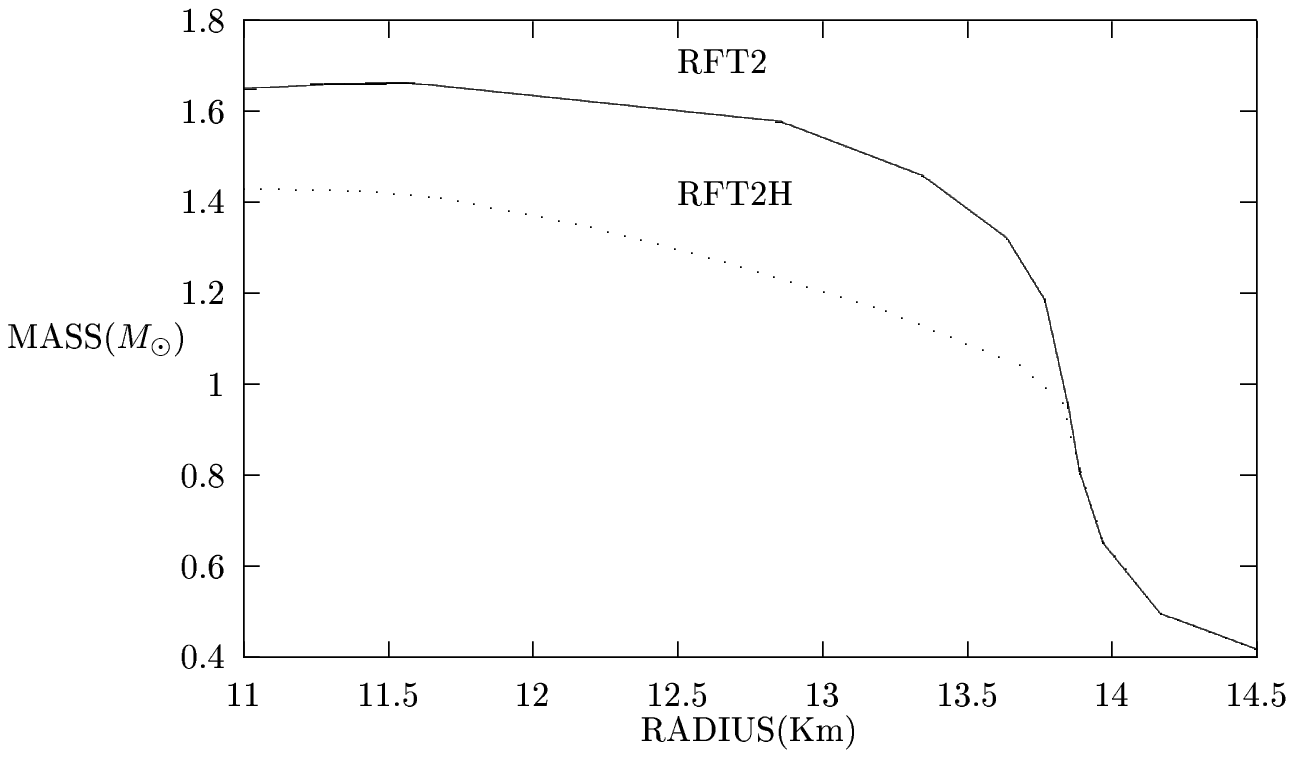,width=2.4in}
\epsfig{file=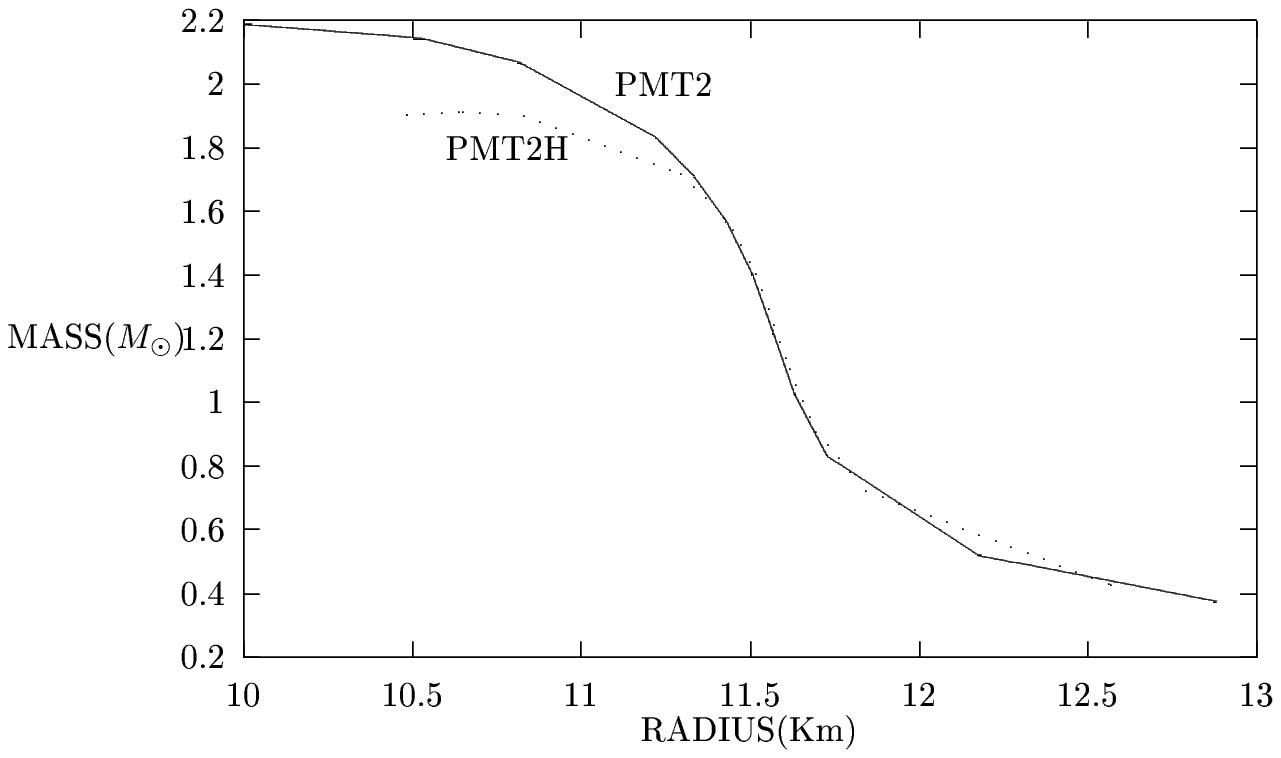,width=2.4in}
\caption{Plot of mass in solar mass unit vs radius in km}
\vskip -0.3cm
\end{figure}
\vskip -.7cm


\end{document}